\documentclass{llncs}
\usepackage[T1]{fontenc}
\usepackage{ae}
\usepackage[english]{babel}
\usepackage[psamsfonts]{amsfonts}
\usepackage{amsfonts,amssymb,stmaryrd,amsmath}
\usepackage{color} 
\usepackage{epsfig}
\usepackage{supertabular}

\def\mi#1{\mathit{#1}}
\def\mb#1{\mathbf{#1}}
\def\mc#1{\mathcal{#1}}
\def\R{\mathbb{R}}
\def\Q{\mathbb{Q}}
\def\Z{\mathbb{Z}}
\def\T{\mathbb{T}}
\def\I{\mathbb{I}}
\def\E{\mathbb{E}}
\def\V{\mc{V}}
\def\Vf{\mc{V}_{\mb{f}}}
\def\F{\mb{F}}
\def\f{\mb{f}}
\def\r{\mb{r}}
\def\fa{\mb{fa}}
\def\mf{\mi{mf}}
\def\Mf{\mi{Mf}}
\def\FF{\mathbb{F}}
\def\FB{\overline{\FF}}
\def\EB{\overline{\E}}
\def\p{\mb{p}}
\def\bias{\mb{bias}}
\def\e{\mb{e}}
\def\O{\mathrm{\Omega}}
\def\expr{\mi{expr}}
\def\const{\mi{const}}
\def\cast{\mi{cast}}
\def\lb#1{\llbracket#1\rrbracket}
\def\lp#1{\llparenthesis\,#1\,\rrparenthesis}
\def\s{\sharp}
\def\ep{\varepsilon_\f}
\def\o{\mb{o}}
\def\widen{\mathbin{\triangledown}}
\def\narrow{\mathbin{\vartriangle}}

\newenvironment{mycenter}{\vspace*{-6pt}\begin{center}}
                         {\end{center}\vspace*{-3pt}}

\newenvironment{mylist}%
        {\begin{list}{$\bullet$}%
        {\setlength{\leftmargin}{1cm}%
         \setlength{\topsep}{0cm}}}%
        {\end{list}}

\makeatletter
\renewcommand\section{\@startsection{section}{1}{0pt}%
                       {-12pt plus -2pt minus -4pt}%
                       {4pt plus 2pt minus 2pt}%
                       {\normalfont\large\bfseries\boldmath
                        \rightskip=0pt plus 8em\pretolerance=10000 }}
\renewcommand\subsection{\@startsection{subsection}{2}{0pt}%
                       {-12pt plus -2pt minus -4pt}%
                       {4pt plus 2pt minus 2pt}%
                       {\normalfont\normalsize\bfseries\boldmath
                        \rightskip=0pt plus 8em\pretolerance=10000 }}
\renewcommand\subsubsection{\@startsection{subsubsection}{3}{0pt}%
                       {-8pt plus -2pt minus -4pt}%
                       {-0.5em plus -0.22em minus -0.1em}%
                       {\normalfont\normalsize\bfseries\boldmath}}
\makeatother

\begin{document}

\newtheorem{thm}{Theorem}


\title{Relational Abstract Domains for the\\
Detection of Floating-Point Run-Time Errors\thanks{This work was
partially supported by the ASTR\'EE RNTL project.}}

\author{Antoine Min\'e}
\institute{DI-\'Ecole Normale Sup\'erieure de Paris, France,\\
           \email{mine@di.ens.fr}}

\maketitle

\begin{abstract}
We present a new idea to adapt relational abstract domains to the analysis of
IEEE 754-compliant floating-point numbers in order to statically detect,
through Abstract Interpretation-based static analyses, potential floating-point
run-time exceptions such as overflows or invalid operations.
In order to take the non-linearity of rounding into account, expressions are
modeled as linear forms with interval coefficients.
We show how to extend already existing numerical abstract domains, such as the
octagon abstract domain, to efficiently abstract transfer functions based on
interval linear forms.
We discuss specific fixpoint stabilization techniques and give some
experimental results.
\end{abstract}


\section{Introduction}

It is a well-established fact, since the failure of the Ariane 5 launcher in 
1996, that run-time errors in critical embedded software can cause great
financial---and human---losses.
Nowadays, embedded software is becoming more and more complex.
One particular trend is to abandon fixed-point arithmetics in favor
of floating-point computations.
Unfortunately, floating-point models are quite complex and features such as
rounding and special numbers (infinities, $\mi{NaN}$, etc.) are not always
understood by programmers.
This has already led to catastrophic behaviors, such as the Patriot
missile story told in \cite{patriot}.

Much work is concerned about the {\it precision} of the computations, that is
to say, characterizing the amount and cause of drift between a
computation on perfect reals and the corresponding floating-point
implementation.
Ours is {\it not}.
We seek to prove that an exceptional behavior (such as division by zero or
overflow) will not occur in any execution of the analyzed program.
While this is a simpler problem, our goal is to scale up to programs of
hundreds of thousands of lines with full data coverage and very few
(even none) false alarms.

Our framework is that of Abstract Interpretation, a generic framework for
designing sound static analyses that already features many instances
\cite{ai,ai2}.
We adapt existing relational numerical abstract domains (generally designed
for the analysis of integer arithmetics) to cope with floating-point
arithmetics.
The need for such domains appeared during the successful design of a
commissioned special-purpose prototype analyzer for a critical
embedded avionics system.
Interval analysis, used in a first prototype \cite{magic1}, proved too
coarse because error-freeness of the analyzed code depends on tests
that are inherently poorly abstracted in non-relational abstract domains.
We also had to design special-purpose widenings and narrowings to
compensate for the pervasive rounding errors, not only in the analyzed program,
but also introduced by our efficient abstractions.
These techniques were implemented in our second prototype whose overall
design was presented in \cite{magic2}.
The present paper focuses on improvements and novel unpublished ideas; it is
also more generic.

\section{Related Work}

\subsubsection{Abstract Domains.}
A key component in Abstract-Interpretation-based analyses is the abstract
domain which is a computer-representable class of program invariants together
with some operators to manipulate them: transfer functions for guards and
assignments, a control-flow join operator, and fixpoint acceleration
operators (such as widenings $\widen$ and narrowings $\narrow$) aiming at
the correct and efficient analysis of loops.
One of the simplest yet useful abstract domain is the widespread interval
domain \cite{ai}.
Relational domains, which are more precise, include Cousot and Halbwachs's
polyhedron domain \cite{poly} (corresponding to invariants of the form
$\sum c_i v_i\leq c$), Min\'e's octagon domain \cite{octagons}
($\pm v_i \pm v_j\leq c$), and Simon's two variables per inequality domain
\cite{tvpli} ($\alpha v_i + \beta v_j \leq c$).
Even though the underlying algorithms for these relational domains
allow them to abstract sets of reals as well as sets of integers, their
efficient implementation---in a maybe approximate but sound way---using
floating-point numbers remains a challenge.
Moreover, these relational domains do not support abstracting floating-point
expressions, but only expressions on perfect integers, rationals, or 
reals.

\subsubsection{Floating-Point Analyses.}
Much work on floating-point is dedicated to the analysis of the precision of 
the computations and the origins of the rounding errors.
The CESTAC method \cite{cestac} is widely used, but also much debated as it
is based on a probabilistic model of error distribution and thus cannot give
sound answers.
An interval-based Abstract Interpretation for error terms is proposed in
\cite{onera}.
Some authors \cite{goubault,martel} go one step further by allowing error
terms to be related in relational, even non-linear, domains.
Unfortunately, this extra precision does not help when analyzing programs
whose correctness also depends upon relations between variables and not only
error terms (such as programs with inequality tests, as in 
Fig.~\ref{limiter}).

\subsubsection{Our Work.}
We first present our IEEE 754-based computation model (Sect.~3) and
recall the classical interval analysis adapted to floating-point
numbers (Sect.~4).
We present, in Sect.~5, an abstraction of floating-point expressions in
terms of interval linear forms over the real field and use it to refine the
interval domain.
Sect.~6 shows how some relational abstract domains can be efficiently
adapted to work on these linear forms.
Sect.~7 presents adapted widening and narrowing techniques.
Finally, some experimental results are shown in Sect.~8.

\section{IEEE 754-Based Floating-Point Model}

We present in this section the concrete floating-point arithmetics model that
we wish to analyze and which is based on the widespread IEEE 754-1985 
\cite{ieee754} norm.

\subsection{IEEE 754 Floating-Point Numbers}

The binary representation of a IEEE 754 number is composed of three
fields:
\begin{mylist}
\item a 1-bit sign $s$;
\item an exponent $e-\bias$, represented by a biased $\e$-bit
unsigned integer $e$;
\item a fraction $f=.b_1\ldots b_\p$, represented by a $\p$-bit
unsigned integer.
\end{mylist}

The values $\e$, $\bias$, and $\p$ are format-specific.
We will denote by $\F$ the set of all available formats and
by $\f=\mb{32}$ the 32-bit {\it single} format
($\e=8$, $\bias=127$, and $\p=23$).
Floating-point numbers belong to one of the following categories:
\begin{mylist}
\item {\it normalized} numbers $(-1)^s\times 2^{e-\bias}\times
1.f$, when $1\leq e\leq 2^\e-2$;
\item {\it denormalized} numbers $(-1)^s\times 2^{1-\bias}\times
0.f$, when $e=0$ and $f\neq 0$;
\item $+0$ or $-0$ (depending on $s$), when $e=0$ and $f=0$;
\item $+\infty$ or $-\infty$ (depending on $s$), when $e=2^\e-1$ and
$f=0$;
\item error codes (so-called {\it NaN}), when $e=2^\e-1$ and $f\neq 0$.
\end{mylist}

For each format $\f\in\F$ we define in particular:
\begin{mylist}
\item $\mf_\f=2^{1-\bias-\p}$ the smallest non-zero positive number;
\item $\Mf_\f=(2-2^{-\p})2^{2^\e-\bias-2}$, the largest non-infinity
number.
\end{mylist}

The special values $+\infty$ and $-\infty$
may be generated as a result of operations
undefined on $\R$ (such as $1/{+0}$), or when a result's absolute
value overflows $\Mf_\f$.
Other undefined operations (such as ${+0}/{+0}$) result in a {\it NaN}
(that stands for {\it Not A Number}).
The sign of $0$ serves only to distinguish between
$1/{+0}=+\infty$ and $1/{-0}=-\infty$; $+0$ and $-0$ are indistinguishable
in all other contexts (even comparison).

Due to the limited number of digits, the result of a floating-point
operation needs to be rounded.
IEEE 754 provides four rounding modes: towards $0$, towards $+\infty$,
towards $-\infty$, and to nearest.
Depending on this mode, either the floating-point number directly smaller or 
directly larger than the exact real result is chosen 
(possibly $+\infty$ or $-\infty$).
Rounding can build infinities from non-infinities operands
(this is called {\it overflow}), and it may return zero when the absolute
value of the result is too small (this is called {\it underflow}).
Because of this rounding phase, most algebraic properties of $\R$, such
as associativity and distributivity, are lost.
However, the opposite of a number is always exactly represented
(unlike what happens in two-complement integer arithmetics), and
comparison operators are also exact.
See \cite{goldberg} for a description of the classical properties and
pitfalls of the floating-point arithmetics.

\subsection{Custom Floating-Point Computation Model}

\begin{figure}[t]
\begin{mycenter}\fbox{
$\begin{array}{lll}

R_{\f,+\infty}(x)&=&\left\{\begin{array}{ll}
\O & \quad\text{if $x> \Mf_\f$}\\
\min\{y\in\FF_\f\;|\;y\geq x\} & \quad\text{otherwise}\\
\end{array}\right.
\\\\

R_{\f,-\infty}(x)&=&\left\{\begin{array}{ll}
\O & \quad\text{if $x< -\Mf_\f$}\\
\max\{y\in\FF_\f\;|\;y\leq x\} & \quad\text{otherwise}\\
\end{array}\right.
\\\\

R_{\f,0}(x)&=&\left\{\begin{array}{ll}
\min\{y\in\FF_\f\;|\;y\geq x\} & \quad\text{if $x\leq 0$}\\
\max\{y\in\FF_\f\;|\;y\leq x\} & \quad\text{if $x\geq 0$}\\
\end{array}\right.
\\\\

R_{\f,n}(x)&=&\left\{\begin{array}{ll}
\O &
\quad\text{if $|x|\geq (2-2^{-\mb{p}-1})2^{2^{\mb{e}}-\mb{bias}-2}$}\\
 \Mf_\f & \quad\text{else if $x\geq  \Mf_\f$}\\
-\Mf_\f & \quad\text{else if $x\leq -\Mf_\f$}\\
R_{\f,-\infty}(x)  & \quad\text{else if
$|R_{\f,-\infty}(x)-x|<|R_{\f,+\infty}(x)-x|$}\\
R_{\f,+\infty}(x)  & \quad\text{else if
$|R_{\f,+\infty}(x)-x|<|R_{\f,-\infty}(x)-x|$}\\
R_{\f,-\infty}(x)  & \quad\text{else if
$R_{\f,-\infty}(x)$'s least significant bit is $0$}\\
R_{\f,+\infty}(x)  & \quad\text{else if
$R_{\f,+\infty}(x)$'s least significant bit is $0$}\\

\end{array}\right.
\end{array}$
}\end{mycenter}
\caption{Rounding functions, extracted from \cite{ieee754}.}
\label{ieeeround}
\end{figure}

\begin{figure}[t]
\begin{mycenter}\fbox{
$\begin{array}{llll}

\lb{\const_{\f,\r}(c)}\rho
&=& R_{\f,\r}(c)\\

\lb{v}\rho &=& \rho(v)\\

\lb{\cast_{\f,\r}(e)}\rho
&=& R_{\f,\r}(\lb{e}\rho)
& \quad\text{if $\lb{e}\rho\neq\O$}\\

\lb{e_1\odot_{\f,\r}e_2}\rho
&=& R_{\f,\r}((\lb{e_1}\rho) \cdot
(\lb{e_2}\rho))
& \quad\text{if $\lb{e_1}\rho,
\lb{e_2}\rho\neq\O,\;
\cdot\in\{+,-,\times\}$}\\

\lb{e_1\oslash_{\f,\r}e_2}\rho
&=& R_{\f,\r}((\lb{e_1}\rho) \slash
(\lb{e_2}\rho))
& \quad\text{if $\lb{e_1}\rho\neq \O,\;
\lb{e_2}\rho\notin\{0,\O\}$}\\

\lb{\ominus e}\rho
&=& -(\lb{e}\rho)
& \quad\text{if $\lb{e}\rho\neq\O$}\\

\lb{\expr_\f}\rho
&=& \O
& \quad\text{in all other cases}\\

\end{array}$
}\end{mycenter}
\caption{Expression concrete semantics, extracted from \cite{ieee754}.}
\label{ieeesem}
\end{figure}

We focus our analysis on the large class of programs that treat floating-point
arithmetics as a practical approximation to the mathematical reals
$\R$: roundings and underflows are tolerated, but not overflows, divisions by
zero or invalid operations, which are considered run-time errors and halt
the program.
Our goal is to detect such behaviors.
In this context, $+\infty$, $-\infty$, and {\it NaN}\/s can never be
created and, as a consequence, the difference between $+0$ and $-0$
becomes irrelevant.
For every format $\f\in\F$, the set of floating-point numbers will be
assimilated to a finite subset of $\R$ denoted by $\FF_\f$.
The grammar of floating-point expressions of format $\f$ includes
constants, variables $v\in\Vf$ of format $\f$, casts (conversion from
another format), binary and unary arithmetic operators (circled in order
to distinguish them from the corresponding operators on reals):
\begin{mycenter}$\begin{array}{lcll}
\expr_\f & :== & \const_{\f,\r}(c) & \quad c\in\R\\
& | & v & \quad v\in\Vf\\
& | & \cast_{\f,\r}(\expr_{\f'}) \\
& | & \expr_\f \odot_{\f,\r} \expr_\f & \quad 
\odot\in\{\oplus,\ominus,\otimes,\oslash\}\\
& | & \ominus \expr_\f
\end{array}$\end{mycenter}

Some constructs are tagged with a floating-point format $\f\in\F$ 
and a rounding
mode $\r\in\{n,0,+\infty,-\infty\}$ ($n$ representing rounding to
nearest).
A notable exception is the unary minus $\ominus$ which does not incur 
rounding and never results in a run-time error as all the $\FF_\f$s are
perfectly symmetric.

An environment $\rho\in\prod_{\f\in\F}(\V_\f\rightarrow\FF_\f)$ is
a function that associates to each variable a floating-point value of
the corresponding format.
Fig.~\ref{ieeesem} gives the concrete semantics
$\lb{\expr_\f}\rho\in\FF_\f\cup\{\O\}$ of the expression $\expr_\f$ in
the environment $\rho$: it can be a number or the run-time error $\O$.
This semantics uses solely the regular operators $+$,$-$,$\times$,$\slash$
on real numbers and the rounding function 
$R_{\f,\r}:\R\rightarrow\FF_\f\cup\{\O\}$ defined in Fig.~\ref{ieeeround}.
It corresponds exactly to the IEEE 754 norm \cite{ieee754} where the overflow,
division by zero, and invalid operation exception traps abort the system
with a run-time error.

\section{Floating-Point Interval Analysis}

\subsubsection{Floating-Point Interval Arithmetics.}
The idea of interval arithmetics is to over-approximate a set of numbers by
an interval represented by its lower and upper bounds.
For each format $\f$, we will denote by $\I_\f$ the set of real intervals
with bounds in $\FF_\f$.
As $\FF_\f$ is totally ordered and bounded, any subset of $\FF_\f$ can be
abstracted by an element of $\I_\f$.
Moreover, as {\it all rounding functions $R_{\f,\r}$ are monotonic}, we
can compute the bounds of any expression using pointing-point operations only,
ensuring efficient implementation within an analyzer.
A first idea is to use the very same format and rounding mode in the
abstract as in the concrete, which would give, for instance, the
following addition ($\O$ denoting a run-time error):
\begin{mycenter}$
{}[a^-;a^+]\oplus^\s_{\f,\r}[b^-;b^+]=\left\{\begin{array}{l}
\O \quad \text{if $a^-\oplus_{\f,\r}b^-=\O$ or $a^+\oplus_{\f,\r}b^+=\O$}\\
{}[a^-\oplus_{\f,\r}b^-;a^+\oplus_{\f,\r}b^+] \quad \text{otherwise}
\end{array}\right.
$\end{mycenter}

A drawback of this semantics is that it requires the analyzer to determine,
for each instruction, which rounding mode $\r$ and which format $\f$ are used.
This may be difficult as the rounding mode can be changed at run-time by a
system call, and compilers are authorized to perform parts of computations
using a more precise IEEE format that what is required by the
programmer (on Intel x86, all floating-point registers are 80-bit wide and
rounding to the user-specified format occurs only when the results are stored
into memory).
Unfortunately, using in the abstract a floating-point format different
from the one used in the concrete computation is not sound.

The following semantics, inspired from the one presented in \cite{goubault},
solves these problems by providing an approximation that is independent
from the concrete rounding mode (assuming always the worst: towards $-\infty$
for the lower bound and towards $+\infty$ for the upper bound):
\begin{mycenter}\begin{supertabular}{ll}
$\bullet$ &
$\const^\s_\f(c) = [\const_{\f,-\infty}(c);\const_{\f,+\infty}(c)]$\\
$\bullet$ & 
$\cast^\s_\f(c) = [\cast_{\f,-\infty}(c);\cast_{\f,+\infty}(c)]$\\
$\bullet$ & 
$[a^-;a^+]\oplus^\s_\f[b^-;b^+]=
{}[a^-\oplus_{\f,-\infty}b^-;a^+\oplus_{\f,+\infty}b^+]$\\
$\bullet$ & 
$[a^-;a^+]\ominus^\s_\f[b^-;b^+]=
{}[a^-\ominus_{\f,-\infty}b^+;a^+\ominus_{\f,+\infty}b^-]$\\
$\bullet$ & 
$[a^-;a^+]\otimes^\s_\f[b^-;b^+]=$\\
& $\quad [\min((a^+\otimes_{\f,-\infty}b^+),(a^-\otimes_{\f,-\infty}b^+),
              (a^+\otimes_{\f,-\infty}b^-),(a^-\otimes_{\f,-\infty}b^-));$\\
& $\quad \;\max((a^+\otimes_{\f,+\infty}b^+),(a^-\otimes_{\f,+\infty}b^+),
               (a^+\otimes_{\f,+\infty}b^-),(a^-\otimes_{\f,+\infty}b^-))]$\\
$\bullet$ & $[a^-;a^+]\oslash^\s_\f[b^-;b^+]=$\\
& $\quad \; \circ \; \O \quad \text{if $b^-\leq 0 \leq b^+$}$\\
& $\quad \; \circ \;
{}[\min((a^+\oslash_{\f,-\infty}b^+),(a^-\oslash_{\f,-\infty}b^+),
 (a^+\oslash_{\f,-\infty}b^-),(a^-\oslash_{\f,-\infty}b^-));$\\
& $\quad \quad
\;\max((a^+\oslash_{\f,+\infty}b^+),(a^-\oslash_{\f,+\infty}b^+),
       (a^+\oslash_{\f,+\infty}b^-),(a^-\oslash_{\f,+\infty}b^-))]$\\
& $\quad \quad \text{otherwise}$\\
$\bullet$ & $\ominus^\s [a^-;a^+]=[\ominus a^+;\ominus a^-]$\\
$\bullet$ & return $\O$ if one interval bound evaluates to $\O$\\
\end{supertabular}\end{mycenter}

This semantics frees the analyzer from the job of statically determining
the rounding mode of the expressions and allows the analyzer to use,
in the abstract, less precise formats that those used in the concrete
(however, using a more precise format in the abstract remains unsound).

\subsubsection{Floating-Point Interval Analysis.}
Interval analysis is a non-relational Abstract Interpretation-based analysis
where, at each program point and for each variable, the set of its possible
values during all executions reaching this point is over-approximated by
an interval. 
An abstract environment $\rho^\s$ is a function
mapping each variable $v\in\V_\f$ to an element of $\I_\f$.
The abstract value $\lb{\expr_\f}^\s\rho^\s\in\I_\f\cup\{\O\}$ of an
expression $\expr_\f$ in an abstract environment $\rho^\s$ can be
derived by induction using the interval operators defined in the preceding
paragraph.

\begin{figure}[t]
\begin{mycenter}\fbox{
\begin{tabular}{l}
{\bf for} (n=0;n<N;n++) \{\\
\quad // fetch $X$ in $[-128;128]$ and $D$ in $[0;16]$\\
\quad $S=Y$; \quad $R=X \ominus_{\mb{32},n} S$; \quad $Y=X$;\\
\quad {\bf if} ($R\leq \ominus D$) $Y=S\ominus_{\mb{32},n}D$;\\
\quad {\bf if} ($R\geq D$) $Y=S\oplus_{\mb{32},n}D$;\\
\}\\
\end{tabular}
}\end{mycenter}
\caption{Simple rate limiter function with input $X$, output $Y$, and
maximal rate variation $D$. }
\label{limiter}
\end{figure}

An assignment $v\leftarrow \expr_\f$ performed in an 
environment $\rho^\s$ returns
$\rho^\s$ where $v$'s value has been replaced by
$\lb{\expr_\f}^\s\rho^\s$ if it does not evaluate to 
$\O$, and otherwise by $\FF_\f$ (the {\it top} value) and reports
an error.
Most tests can only be abstracted by ignoring them (which is sound).
Even though for simple tests such as, for instance, $X\leq Y\oplus_{\f,\r}c$, 
the
interval domain is able to refine the bounds for $X$ and $Y$, it cannot
remember the relationship between these variables.
Consider the more complete example of Fig.~4.
It is a rate limiter that given random input flows $X$ and $D$, bounded
respectively by $[-128;128]$ and $[0;16]$, computes an output flow $Y$
that tries to follow $X$ while having a change rate limited by $D$.
Due to the imprecise abstraction of tests, the interval domain will bound 
$Y$ by $[-128-16n;128+16n]$ after $n$ loop iterations while in fact it is 
bounded by $[-128;128]$ independently from $n$.
If $N$ is too big, the interval analysis will conclude that the limiter
may overflow while it is in fact always perfectly safe.

\section{Linearization of Floating-Point Expressions}

Unlike the interval domain, relational abstract domains rely on 
algebraic properties of operators, such as associativity and distributivity, 
that are not true in the floating-point world.
Our solution is to approximate floating-point expressions by
{\it linear expressions} in the {\it real} field with 
{\it interval coefficients} and free variables in
$\V=\cup_{\f\in\F}\V_\f$.
Let $i+\sum_{v\in\V}i_v v$ be such a linear form; it can be viewed as
a function from $\R^\V$ to the set of real intervals.
For the sake of efficiency, interval coefficient bounds will be
represented by floating-point numbers in a format $\fa$ that is efficient
on the analyzer's platform: $i,i_v\in\I_\fa$.
Because, for all $\f$ and $\cdot\in\{+,-,\times,\slash\}$,
$\odot^\s_\f$ is a valid over-approximation of the
corresponding real interval arithmetics operation, we can define the following
sound operators $\boxplus^\s$, $\boxminus^\s$, $\boxtimes^\s$, $\boxslash^\s$
on linear forms:
\begin{mycenter}\begin{supertabular}{llll}
$\bullet$ & $(i+\sum_{v\in\V}i_vv)\boxplus^\s(i'+\sum_{v\in\V}i'_vv)$
&=& $(i\oplus^\s_\fa i')+\sum_{v\in\V}(i_v\oplus^\s_\fa i'_v)v$\\

$\bullet$ & $(i+\sum_{v\in\V}i_vv)\boxminus^\s(i'+\sum_{v\in\V}i'_vv)$
&=& $(i\ominus^\s_\fa i')+\sum_{v\in\V}(i_v\ominus^\s_\fa i'_v)v$\\

$\bullet$ & $i\boxtimes^\s(i'+\sum_{v\in\V}i'_vv)$
&=& $(i\otimes^\s_\fa i')+\sum_{v\in\V}(i\otimes^\s_\fa i'_v)v$\\

$\bullet$ & $(i+\sum_{v\in\V}i_vv)\boxslash^\s i'$
&=& $(i\oslash^\s_\fa i')+\sum_{v\in\V}(i_v\oslash^\s_\fa i')v$\\
\end{supertabular}\end{mycenter}

Given an expression $\expr_\f$ and an interval abstract environment
$\rho^\s$ as in Sect.~4, we construct the interval linear form 
$\lp{\expr_\f}\rho^\s$ on $\V$ as follows:
\begin{mycenter}$\begin{array}{llll}
\bullet & \lp{\const_{\f,\r}(c)}\rho^\s & = &
[\const_{\f,-\infty}(c);\const_{\f,+\infty}(c)]\\

\bullet & \lp{v_\f}\rho^\s & = & [1;1]v_\f\\

\bullet & \lp{\cast_{\f,\r}(e)}\rho^\s & = &
\lp{e}\rho^\s \;\boxplus^\s\; \ep(\lp{e}\rho^\s ) \;\boxplus^\s\; 
\mf_\f[-1;1]\\

\bullet & \lp{e_1\oplus_{\f,\r}e_2}\rho^\s & = & \\
\multicolumn{4}{l}{\quad
\lp{e_1}\rho^\s  \;\boxplus^\s\; \lp{e_2}\rho^\s \;\boxplus^\s\; 
\ep(\lp{e_1}\rho^\s)  \;\boxplus^\s\; \ep(\lp{e_2}\rho^\s)
\;\boxplus^\s\;  \mf_\f[-1;1]}\\

\bullet & \lp{e_1\ominus_{\f,\r}e_2}\rho^\s & = & \\
\multicolumn{4}{l}{\quad
\lp{e_1}\rho^\s  \;\boxminus^\s\; \lp{e_2}\rho^\s \;\boxplus^\s\; 
\ep(\lp{e_1}\rho^\s)  \;\boxplus^\s\; \ep(\lp{e_2}\rho^\s)
\;\boxplus^\s\;  \mf_\f[-1;1]}\\

\bullet & \lp{[a;b]\otimes_{\f,\r}e_2}\rho^\s & = & \\
\multicolumn{4}{l}{\quad
([a;b]\boxtimes^\s\lp{e_2}\rho^\s) \;\boxplus^\s\; 
([a;b]\boxtimes^\s\ep(\lp{e_2}\rho^\s))
\;\boxplus^\s\;  \mf_\f[-1;1]}\\

\bullet & \lp{e_1\otimes_{\f,\r}[a;b]}\rho^\s & = & 
\lp{[a;b]\otimes_{\f,\r}e_1}\rho^\s \\

\bullet & \lp{e_1\otimes_{\f,\r}e_2}\rho^\s & = & 
\lp{\iota(\lp{e_1}\rho^\s)\rho^\s\;\otimes_{\f,\r}\;e_2}\rho^\s \\

\bullet & \lp{e_1\oslash_{\f,\r}[a;b]}\rho^\s & = & \\
\multicolumn{4}{l}{\quad
(\lp{e_1}\rho^\s\boxslash^\s[a;b]) \;\boxplus^\s\; 
(\ep(\lp{e_1}\rho^\s)\boxslash^\s [a;b])
\;\boxplus^\s\;  \mf_\f[-1;1]} \\

\bullet & \lp{e_1\oslash_{\f,\r}e_2}\rho^\s & = & 
\lp{e_1\;\oslash_{\f,\r}\;\iota(\lp{e_2}\rho^\s)\rho^\s} \\

\end{array}$\end{mycenter}

where the ``error'' $\ep(l)$ of the linear form $l$ is the following
linear form:
\begin{mycenter}$\begin{array}{lll}
\displaystyle
\ep\left([a;b]+\sum_{v\in\V}[a_v;b_v]v\right) &=&
(\max(|a|,|b|)\otimes^\s_\fa[-2^{-\p};2^{-\p}])\;+\\
&&\displaystyle
\sum_{v\in\V}(\max(|a_v|,|b_v|)\otimes^\s_\fa[-2^{-\p};2^{-\p}])v\\
\multicolumn{3}{l}{\text{{\it($\p$ is the fraction size in bits for the format
$\f$, see Sect.~3.1)}}}
\end{array}$\end{mycenter}

and the ``intervalization'' $\iota(l)\rho^\s$ function over-approximates
the range of the linear form $l$ in the abstract environment $\rho^\s$ as the 
following interval of $\I_\fa$:
\begin{mycenter}$\begin{array}{l}
\displaystyle
\iota\left(i+\sum_{v\in\V}i_vv\right)\rho^\s =
i\oplus^\s_\fa\left(\sideset{}{^\s_\fa}\bigoplus_{v\in\V}i_v\otimes^\s_\fa
\rho^\s(v)\right)\\
\text{{\it (any summation order for $\oplus^\s_\fa$ is sound)}}
\end{array}$\end{mycenter}

Note that this semantics is very different from the one proposed by Goubault
in \cite{goubault} and subsequently used in \cite{martel}.
In \cite{goubault}, each operation introduces a new variable representing an
error term, and there is no need for interval coefficients.

\subsubsection{About $\iota$.}
Dividing a linear form by another linear form which is not reduced to an
interval does not yield a linear form.
In this case, the $\iota$ operator is used to over-approximate the divisor
by a single interval before performing the division.
The same holds when multiplying two linear forms not reduced to an
interval, but we can choose
to apply $\iota$ to either argument.
For the sake of simplicity, we chose here to ``intervalize'' the left
argument.
Moreover, any non-linear operator (such as, e.g., square root or sine) could
be dealt with by performing the corresponding operator on intervals after
``intervalizing'' its argument(s).

\subsubsection{About $\ep$.}
To account for rounding errors, an upper bound of 
$|R_{\f,\r}(x\cdot y)-(x\cdot y)|$ (where $\cdot\in\{+,-,\times,\slash\}$) is
included in $\lp{\expr_\f}\rho^\s$.
It is the sum an error relative to the arguments $x$ and $y$, expressed
using $\ep$, and an absolute error $\mf_\f$ due to a possible underflow.
Unlike what happened with interval arithmetics, correct error computation
does not require the abstract operators to use floating-point formats
that are no more precise than the concrete ones: the choice of $\fa$ is
completely free.

\subsubsection{About $\O$.}
It is quite possible that, during the computation of 
$\lp{\expr_\f}\rho^\s$, a floating-point run-time error $\O$ occurs.
In that case, we will say that the linearization ``failed''.
It does not mean that the program has a run-time error, but only that
we cannot compute a linearized expression and must revert to the classical
interval arithmetics.

\subsubsection{Main Result.}
When we evaluate in $\R$ the linear form $\lp{\expr_\f}\rho^\s$ in
a concrete environment $\rho$ included in $\rho^\s$ we get a real
interval that over-approximates the concrete value of the expression:

\begin{theorem}
~\par\noindent\begin{tabular}{l}
\noindent
If $\lb{\expr_\f}^\s\rho^\s\neq\O$ and the linearization does not fail and
$\forall v\in\V,\;\rho(v)\in\rho^\s(v)$,\\
\noindent
then $\lb{\expr_\f}\rho\in\lp{\expr_\f}\rho^\s\:(\rho)$.
\end{tabular}
\end{theorem}

\subsubsection{Linear Form Propagation.}
As the linearization manipulates expressions symbolically, it is able to
perform simplifications when the same variable appears several times.
For instance $Z\leftarrow X\ominus_{\mb{32},n}(0.25\otimes_{\mb{32},n}X)$
will be interpreted as 
$Z\leftarrow[0.749\cdots;0.750\cdots]X+2.35\cdots 10^{-38}[-1;1]$.
Unfortunately, no simplification can be done if the expression is broken into
several statements, such as in
$Y\leftarrow 0.25\otimes_{\mb{32},n}X; Z\leftarrow X\ominus_{\mb{32},n}Y$.
Our solution is to remember, in an extra environment $\rho^\s_l$,
the linear form assigned to each variable and use this information while
linearizing: we set $\lp{v_\f}(\rho^\s,\rho^\s_l)=\rho^\s_l(v)$ instead
of $[-1;1]v_\f$.
Care must be taken, when a variable $v$ is modified, to discard all occurrences
of $v$ in $\rho^\s_l$.
Effects of tests on $\rho^\s_l$ are ignored.
Our partial order on linear forms is flat, so, at control-flow joins, 
only variables that are associated with the same linear form in both
environments are kept; moreover, we do not need any widening.
This technique is reminiscent of Kildall's constant propagation \cite{kildall}.

\subsubsection{Applications.}
A first application of linearization is to improve the precision of the
interval analysis.
We simply replace $\lb{\expr_\f}^\s\rho^\s$ by
$\lb{\expr_\f}^\s\rho^\s\;\cap\;\iota(\lp{\expr_\f}\rho^\s)\rho^\s$ 
whenever the hypotheses of Thm.~1 hold.

While improving the assignment transfer function (trough expression
simplification), this is not sufficient to treat tests precisely.
For this, we need relational domains.
Fortunately, Thm.~1 also means that if we have a relational domain that
manipulates sets of points with real coordinates $\R^\V$ and that is
able to perform assignments and tests of linear expressions with interval
coefficients, we can use it to perform relational analyses on floating-point
variables.
Consider, for instance, the following algorithm to handle an assignment
$v\leftarrow \expr_\f$ in such a relational domain (the procedure
would be equivalent for tests):
\begin{mylist}
\item If $\lb{\expr_\f}^\s\rho^\s=\O$, then we report a run-time error and 
apply the transfer function for $v\leftarrow\FF_\f$.
\item Else, if the linearization of $\expr_\f$ fails, then we do not report
an error but apply the transfer function for 
$v\leftarrow\lb{\expr_\f}^\s\rho^\s$.
\item Otherwise, we do not report an error but apply the transfer function for
$v\leftarrow\lp{\expr_\f}\rho^\s$.
\end{mylist}
Remark how we use the interval arithmetics to perform the actual 
detection of run-time errors and as a fallback when the linearization cannot be
used.

\section{Adapting Relational Abstract Domains}

We first present in details the adaptation of the octagon abstract domain 
\cite{octagons} 
to use floating-point arithmetics and interval linear forms, which
was implemented in our second prototype analyzer \cite{magic2}.
We then present in less details some ideas to adapt other domains.

\subsection{The Octagon Abstract Domain}
The octagon abstract domain \cite{octagons} can manipulate sets of constraints
of the form $\pm x\pm y\leq c$, $x,y\in\V$, $c\in\E$ where $\E$ can
be $\Z$, $\Q$, or $\R$.
An abstract element $\o$ is represented by a half-square constraint matrix 
of size $|\V|$.
Each element at line $i$, column $j$ with $i\leq j$ contains four constraints:
$v_i+v_j\leq c$, $v_i-v_j\leq c$, $-v_i+v_j\leq c$, and $-v_i-v_j\leq c$, with
$c\in\EB=\E\cup\{+\infty\}$.
Remark that diagonal elements represent interval constraints as
$2v_i\leq c$ and $-2v_i\leq c$.
In the following, we will use notations such as $\max_\o(v_i+v_j)$ to 
access the upper bound, in $\EB$, of constraints embedded 
in the octagon $\o$.

Because constraints in the matrix can be combined to obtain implied
constraints that may not be in the matrix (e.g., from $x-y\leq c$ and
$y+z\leq d$, we can deduce $x+z\leq c+d$), two matrices can represent the same
set of points.
We introduced in \cite{octagons} a Floyd-Warshall-based closure operator that
provides a normal form by combining and propagating, in $\mc{O}(|\V|^3)$
time, all constraints.
The optimality of the abstract operators requires to work on closed matrices.

\subsubsection{Floating-Point Octagons.}
In order to represent and manipulate efficiently constraints on real numbers,
we choose to use floating-point matrices: $\FF_\fa$ replaces $\E$
(where $\fa$ is, as before, an efficient floating-point format chosen by the 
analyzer implementation).
As the algorithms presented in \cite{octagons} make solely use of the $+$ and 
$\leq$ operators on $\EB$, it is sufficient to replace
$+$ by $\oplus_{\fa,+\infty}$ and map $\O$ to $+\infty$ in these algorithms to
provide a sound approximation of all the transfer functions and operators
on reals using only $\FB_\fa=\FF_\fa\cup\{+\infty\}$.
As all the nice properties of $\E$ are no longer true in $\FF_\fa$, the closure
is no longer a normal form.
Even though working on closed matrices will no longer guaranty the optimality
of the transfer functions, it still greatly improves their precision.

\subsubsection{Assignments.}
Given an assignment of the form 
$v_k\leftarrow l$, where $l$ is a interval linear form, on an octagon $\o$, the
resulting set is no always an octagon.
We can choose between several levels of approximation.
Optimality could be achieved at great cost by performing the assignment
in a polyhedron domain and then computing the smallest enclosing octagon.
We propose, as a less precise but much faster alternative
($\mc{O}(|\V|$ time cost), to replace
all constraints concerning the variable $v_k$ by the following ones:
\begin{mycenter}$\begin{array}{llll}
v_k+v_i & \leq & u(\o,l\boxplus^\s v_i) & \forall i\neq k\\
v_k-v_i & \leq & u(\o,l\boxminus^\s v_i) & \forall i\neq k\\
-v_k+v_i & \leq & u(\o,v_i\boxminus^\s l) & \forall i\neq k\\
-v_k-v_i & \leq & u(\o,\boxminus^\s(l\boxplus^\s v_i)) \quad 
& \forall i\neq k\\
v_k & \leq & u(\o,l) \\
-v_k & \leq & u(\o,\boxminus^\s l) \\
\end{array}$\end{mycenter}

where the upper bound of a linear form $l$ on an octagon $\o$ is approximated
by $u(\o,l)\in\FB_\fa$ as follows:
\begin{mycenter}$\begin{array}{l}\displaystyle
u \left(\o,\;[a^-;a^+]+\sum_{v\in\V} [a^-_v;a^+_v] v\right) =
a^+ \;\oplus_{\fa,+\infty}
\vspace*{-8pt}
\\\displaystyle
\qquad\quad
\left(
\sideset{}{_{\fa,+\infty}}\bigoplus_{v\in\V}
\max (\begin{array}[t]{l}
 \max_\o (v) \otimes_{\fa,+\infty} a^+_v,\;
 \ominus(\max_\o (-v) \otimes_{\fa,-\infty} a^+_v),\\
 \max_\o (v) \otimes_{\fa,+\infty} a^-_v,\;
 \ominus(\max_\o (-v) \otimes_{\fa,-\infty} a^-_v)
)\end{array}
\right)
\vspace*{6pt}\\
\text{{\it (any summation order for $\oplus_{\fa,+\infty}$ is sound)}}
\end{array}
$\end{mycenter}
and $\oplus_{\fa,+\infty}$ and 
$\otimes_{\fa,+\infty}$ are extended to $\FB_\fa$ as follows:
\begin{mycenter}$\begin{array}{l}
+\infty\oplus_{\fa,+\infty}x = x\oplus_{\fa,+\infty}+\infty=+\infty\\
+\infty\otimes_{\fa,+\infty}x = x\otimes_{\fa,+\infty}+\infty=
\left\{\begin{array}{ll}
0&\quad\text{if $x=0$}\\
+\infty&\quad\text{otherwise}
\end{array}\right.
\end{array}$\end{mycenter}

\begin{example}
\label{assign}
Consider the assignment $X=Y\oplus_{\mb{32},n}Z$ with $Y,Z\in[0;1]$.
It is linearized as $X=[1-2^{-23};1+2^{-23}](Y+Z)+\mf_\mb{32}[-1;1]$,
so our abstract transfer function will infer relational constraints 
such as $X-Y\leq 1+2^{-22}+\mf_\mb{32}$.
\end{example}

\subsubsection{Tests.}
Given a test of the form $l_1\leq l_2$, where $l_1$ and $l_2$ are
linear forms, for all variable $v_i\neq v_j$, appearing in 
$l_1$ or $l_2$, the constraints in the octagon $\o$ can be tightened by
adding the following extra constraints:

\begin{mycenter}\begin{supertabular}{lll}
$v_j-v_i$  &$\leq$&  
$u(\o,\;l_2\boxminus^\s l_1\boxminus^\s v_i\boxplus^\s v_j)$\\
$v_j+v_i$  &$\leq$& 
$u(\o,\;l_2\boxminus^\s l_1\boxplus^\s v_i\boxplus^\s v_j)$\\
$-v_j-v_i$  &$\leq$&
$u(\o,\;l_2\boxminus^\s l_1\boxminus^\s v_i\boxminus^\s v_j)$\\
$-v_j+v_i$  &$\leq$&
$u(\o,\;l_2\boxminus^\s l_1\boxplus^\s v_i\boxminus^\s v_j)$\\
$v_i$ &$\leq$&  
$u(\o,\;l_2\boxminus^\s l_1\boxplus^\s v_i)$\\
$-v_i$ &$\leq$&  
$u(\o,\;l_2\boxminus^\s l_1\boxminus^\s v_i)$\\
\end{supertabular}\end{mycenter}

\begin{example}
\label{test}
Consider the test $Y\oplus_{\mb{32},n}Z\leq 1$ with $Y,Z\in[0;1]$.
It is linearized as $[1-2^{-23};1+2^{-23}](Y+Z)+\mf_\mb{32}[-1;1]\leq [1;1]$.
Our abstract transfer function will be able to infer the constraint:
$Y+Z\leq 1+2^{-22}+\mf_\mb{32}$.
\end{example}

\begin{example}
The optimal analysis of the rate limiter function of Fig.~\ref{limiter} would
require representing interval linear invariants on {\it three} variables.
Nevertheless, the octagon domain with our approximated transfer functions 
can prove that the output $Y$ is bounded by $[-136;136]$ 
independently from $n$ (the optimal bound being $[-128;128]$), 
which is sufficient to prove that $Y$ does not overflow.
\end{example}

\subsubsection{Reduction with Intervals.}
The interval environment $\rho^\s$ 
is important as we use it to perform run-time error
checking and to compute the linear form associated to an expression.
So, we suppose that transfer functions are performed in parallel in the
interval domain and in the octagon domain, and then, information from
the octagon result $\o$ is used to refine the interval result
$\rho^\s$ as follows:
for each variable $v\in\V$, the upper bound of 
$\rho^\s(v)$ is replaced by
$\min (\max \rho^\s(v),\max_\o(v))$ and the same is done for its lower bound.

\subsection{Polyhedron Domain}

The polyhedron domain is much more precise than the octagon domain as it
allows manipulating sets of invariants of the form $\sum_v c_v v\leq c$,
but it is also much more costly.
Implementations, such as the New Polka or the 
Parma Polyhedra libraries \cite{parma}, are
targeted at representing sets of points with integer or rational coordinates.
They internally use rational coefficients and, as the coefficients usually
become fairly large, arbitrary precision integer libraries.

\subsubsection{Representing Reals.}
These implementations could be used as-is to abstract sets of points with
real coordinates, but the rational coefficients may get out of control, as
well as the time cost.
Unlike what happened for octagons, it is not so easy to adapt the algorithms
to floating-point coefficients while retaining soundness as they are much
much more complex.
We are not aware, at the time of writing, of any such floating-point 
implementation.

\subsubsection{Assignments and Tests.}
Assignments of the form $v\leftarrow l$ and tests of the form $l\leq 0$
where $l=[a^-;a^+] + \sum_{v\in\V}[a^-_v;a^+_v]v$ 
seem difficult to abstract in general.
However, the case where all coefficients in $l$ are scalar except maybe the
constant one is much easier.
To cope with the general case, an idea (yet untested) 
is to use the following transformation 
that abstracts $l$ into an over-approximated linear form $l'$ where
$\forall v,\;a^-_v=a^+_v$ by transforming all relative errors into
absolute ones:
\begin{mycenter}$\begin{array}{lll}
l' &=&\displaystyle
\left(
[a^-;a^+]\oplus^\s_\fa
\sideset{}{^\s_\fa}\bigoplus_{v\in\V} 
(a^+_v \ominus_{\fa,+\infty} a^-_v)\otimes^\s_\fa
[0.5;0.5]\otimes^\s_\fa\rho^\s(v)\right)\;+
\\
&&\displaystyle
\sum_{v\in\V}
((a^-_v \oplus_{\fa,+\infty} a^+_v)\otimes^\s_\fa[0.5;0.5]) v
\vspace*{4pt}\\
\multicolumn{3}{l}{
\text{{\it (any summation order for $\oplus^\s_\fa$ is sound)}}}
\end{array}$\end{mycenter}

\subsection{Two Variables per Linear Inequalities Domain}
Simon's domain \cite{tvpli} can manipulate constraints of the form
$\alpha v_i+\beta v_j\leq c$, $\alpha,\beta,c\in\Q$.
An abstract invariant is represented using a planar convex polyhedron for
each pair of variables.
As for octagons, most computations are done point-wise on variable pairs
and a closure provides the normal form by propagating and combining 
constraints.
Because the underlying algorithms are simpler than for generic polyhedra,
adapting this domain to handle floating-point computations efficiently
may prove easier while greatly improving the precision over octagons.
This still remains an open issue.

\subsection{Ellipsoid and Digital Filter Domains}

During the design of our prototype analyzer \cite{magic2}, we encountered
code for computing recursive sequences such as
 $X_i=((\alpha \otimes_{\mb{32},n} X_{i-1})\oplus_{\mb{32},n}
(\beta \otimes_{\mb{32},n}X_{i-2}))\oplus_{\mb{32},n}\gamma$
(1), or $X_i=(\alpha \otimes_{\mb{32},n}X_{i-1})\oplus_{\mb{32},n}
(Y_i\ominus_{\mb{32},n}Y_{i-1})$ (2).
In order to find precise bounds for the variable $X$, one has to consider
invariants out of the scope of classical relational abstract domains.
Case (1) can be solved by using the ellipsoid abstract domain of
\cite{magic2} that can represent non-linear real invariants of the form
$aX_i^2+bX_{i-1}^2+cX_iX_{i-1}\leq d$, while case (2) is precisely
analyzed using Feret's filter domains \cite{filters} by inferring
temporal invariants of the form $|X_i|\leq a\max_{j\leq i} |Y_j|+b$.
It is not our purpose here to present these new abstract domains
but we stress the fact that such domains, as the ones discussed in the
preceding paragraphs, are naturally designed to work with perfect reals, but
used to analyze imperfect floating-point computations.

A solution is, as before, to design these domains to analyze interval linear
assignments and tests on reals, and feed them with the result of the
linearization of floating-point expressions defined in Sect.~5.
This solution has been successfully applied (see \cite{filters} and
Sect.~8).

\section{Convergence Acceleration}
\label{widen}

In the Abstract Interpretation framework, loop invariants are described
as fixpoints and are over-approximated by iterating, in the abstract,
the body transfer function $F^\s$ until a post-fixpoint is reached.

\subsubsection{Widening.}
The widening $\widen$ is a convergence acceleration operator introduced
in \cite{ai} in order to reduce the number of abstract iterations:
$\lim_i (F^\s)^i$ is replaced by $\lim_i E^\s_i$ where
$E^\s_{i+1}=E^\s_i\widen F^\s(E^\s_i)$.
A straightforward widening on intervals and octagons is to simply discard
unstable constraints.
However, this strategy is too aggressive and fails to discover sequences 
that are stable after a certain bound, such as, e.g.,
$X=(\alpha\otimes_{\mb{32},n}X)\oplus_{\mb{32},n}\beta$.
To give these computations a chance to stabilize, we use a staged widening
that tries a user-supplied set of bounds in increasing order.
As we do not know in advance which bounds will be stable, we use,
as set $\T$ of thresholds, a simple exponential ramp:
$\T=\{\pm 2^i\}\cap\FF_\fa$.
Given two octagons $\o$ and $\o'$, the widening with thresholds
$\o\widen\o'$ is obtained by setting, for each binary unit
expression $\pm v_i\pm v_j$:
\begin{mycenter}$
\max_{\o\widen\o'} (C) =
\left\{\begin{array}{ll}
\max_{\o}(C) & \text{if $\max_{\o'}(C)\leq\max_{\o}(C)$}\\
\min \{t\in\T\cup\{+\infty\}\;|\;t\geq \max_{\o'}(C)\} & 
\text{otherwise}\\
\end{array}\right.
$\end{mycenter}

\subsubsection{Decreasing Iterations.}
We now suppose that we have iterated the widening with thresholds up to an
abstract post-fixpoint $X^\s$: $F^\s(X^\s)\sqsubseteq X^\s$.
The bound of a stable variable is generally over-approximated by the
threshold immediately above.
One solution to improve such a bound is to perform some decreasing iterations
$X^\s_{i+1}=X^\s_i\sqcap F(X^\s_i)$ from
$X^\s_0=X^\s$.
We can stop whenever we wish, the result will always be, by construction, 
an abstraction of the concrete fixpoint;
however, it may no longer be a post-fixpoint for $F^\s$.
It is desirable for invariants to be abstract post-fixpoint so that the
analyzer can check them independently from the way they were generated instead
of relying solely on the maybe buggy fixpoint engine.

\subsubsection{Iteration Perturbation.}
Careful examination of the iterates on our benchmarks
showed that the reason we do not get an abstract post-fixpoint
is that the {\it abstract} computations are done
in floating-point which incurs a somewhat non-deterministic extra rounding.
There exists, between $F^\s$'s definitive pre-fixpoints and
$F^\s$'s definitive post-fixpoints, a chaotic region.
To ensure that the $X^\s_i$ stay above this region, we replace
the intersection $\sqcap$ used in the decreasing iterations 
by the following narrowing $\narrow$:
$\o\narrow\o'=\epsilon(\o\sqcap\o')$
where $\epsilon(\o)$ returns an octagon where the bound of each
unstable constraint is enlarged by $\epsilon \times d$, where $d$ is the 
maximum of all non $+\infty$ constraint bounds in $\o$.
Moreover, replacing $\o\widen\o'$ by $\epsilon(\o\widen\o')$
allows the analyzer to skip above $F^\s$'s chaotic regions
and effectively reduces the required number of increasing iterations, and so,
the analysis time.

Theoretically, a good $\epsilon$ can be estimated by  the relative amount of 
rounding errors performed in the abstract computation of one loop iteration, 
and so, is a function of the complexity of the analyzed loop body, 
the floating-point format $\fa$ used in the analyzer and the
implementation of the abstract domains.
We chose to fix $\epsilon$ experimentally by enlarging a small value
until the analyzer reported it found an abstract post-fixpoint for our program.
Then, as we improved our abstract domains and modified the analyzed program,
we seldom had to adjust this $\epsilon$ value.

\section{Experimental Results}

We now show how the presented abstract domains perform in practice.
Our only real-life example is the critical embedded avionics software of
\cite{magic2}.
It is a $132,000$ lines reactive C program 
(75 KLoc after preprocessing) containing approximately $10,000$ global 
variables, $5,000$ of which are floating-point variables, single precision.
The program consists mostly of one very large loop executed
$3.6\cdot 10^6$ times.
Because relating several thousands variables in a relational domain
is too costly, we use the ``packing'' technique described in \cite{magic2}
to statically determine sets of variables that should be related together
and we end up with approximately $2,400$ octagons of size
$2$ to $42$ instead of one octagon of size $10,000$.

\begin{figure}[t]
\begin{mycenter}
\begin{tabular}{c|c|c|c|c|c|c|c|}
\cline{2-8}
&\multicolumn{3}{|c|}{{\bf domains}}&
&{\bf nb. of}&&{\bf nb. of}\\
\cline{2-4}
&\;{\bf linearize}\;&\;{\bf octagons}\;&\;{\bf filters}\;&
\;\;{\bf time}\;\;&\;{\bf iterations}\;&\;{\bf memory}\;&\;{\bf alarms}\;\\
\cline{2-8}
(1)&
$\times$&$\times$&$\times$&
1623 s & 150 & 115 MB & 922 \\
\cline{2-8}
(2)&
$\surd$&$\times$&$\times$&
4001 s & 176 & 119 MB & 825 \\
\cline{2-8}
(3)&
$\surd$&$\surd$ &$\times$& 
3227 s & 69 & 175 MB & 639 \\
\cline{2-8}
(4)&
$\surd$&$\times$&$\surd$& 
8939 s & 211 & 207 MB & 363 \\
\cline{2-8}
(5)&
$\surd$&$\surd$&$\surd$& 
4541 s & 72 & 263 MB & 6 \\
\cline{2-8}
\end{tabular}
\end{mycenter}
\caption{Experimental results.}
\label{results1}
\end{figure}

Fig.~\ref{results1} shows how the choice of the abstract domains influence
the precision and the cost of the analysis presented in \cite{magic2} on
our 2.8 GHz Intel Xeon.
Together with the computation time, we also give the number of abstract 
executions of the big loop needed to find an invariant; thanks to our widenings
and narrowings, it is much much less than the concrete number of iterations.
All cases use the interval domain with the symbolic simplification
automatically provided by the linearization, except (1) that uses plain 
interval analysis.
Other lines show the influence of the octagon (Sect.~6.1) and the specialized 
digital filter domains (\cite{filters} and Sect.~6.4):
when both are activated, we only get six potential run-time errors
for a reasonable time and memory cost.
This is a sufficiently small number of alarms to allow manual inspection, and
we discovered they could be eliminated without altering the functionality
of the application by changing only three lines of code.
Remark that as we add more complex domains, the time cost per iteration
grows but the number of iterations needed to find an invariant decreases so 
that a better precision may reduce the overall time cost.

\section{Conclusion}

We presented, in this paper, an adaptation of the octagon abstract domain in
order to analyze programs containing IEEE 754-compliant floating-point
operations.
Our methodology is somewhat generic and we proposed some ideas to adapt 
other relational numerical abstract domains as well.
The adapted octagon domain was implemented in our prototype static analyzer
for run-time error checking of critical C code \cite{magic2} and tested
on a real-life embedded avionic application.
Practical results show that the proposed method scales up well and does 
greatly improve the precision of the analysis when compared to the classical
interval abstract domain while maintaining a reasonable cost.
To our knowledge, this is the first time relational numerical domains are used
to represent relations between floating-point variables.

\subsubsection{Acknowledgments.}
We would like to thank all the members of the ``magic'' team: 
Bruno Blanchet, Patrick Cousot, Radhia Cousot, J\'er\^ome
Feret, Laurent Mauborgne, David Monniaux, Xavier Rival, as well as
the anonymous referees.


\bibliographystyle{plain}
\bibliography{bibarticle}

\end{document}